# Modular interface for managing cognitive bias in experts

Melody G. Whitehead[1*] and Andrew Curtis[2]

[1] Institution: Massey University

[1] Address: Volcanic Risk Solutions, Massey University, Private Bag 11222, Palmerston North 4442, New Zealand

[2] Institution: University of Edinburgh

[2] Address: Grant Institute, University of Edinburgh, James Hutton Road, Edinburgh, EH9 3FE, United Kingdom

*Corresponding author: m.whitehead@massey.ac.nz

**Abstract**

Expert knowledge is required to interpret data across a range of fields. Experts bridge gaps that often exists in our knowledge about relationships between data and the parameters of interest. This is especially true in geoscientific applications, where knowledge of the Earth is derived from interpretations of observable features and relies on predominantly unproven but widely accepted theories. Thus, experts facilitate solutions to otherwise unsolvable problems. However, experts are inherently subjective, and susceptible to cognitive biases and adverse external effects. This work examines this problem within geoscience. Three compelling examples are provided of the prevalence of cognitive biases from previous work. The problem is then formally defined, and a set of design principles which ensure that any solution is sufficiently flexible to be readily applied to the range of geoscientific problems. No solutions exist that reliably capture and reduce cognitive bias in experts. However, formal expert elicitation methods can be used to assess expert variation, and a variety of approaches exist that may help to illuminate uncertainties, avoid misunderstandings, and reduce herding behaviours or single-expert over-dominance. This work combines existing and future approaches to reduce expert suboptimality through a flexible modular design where each module provides a specific function. The design centres around action modules that force a stop-and-perform step into interpretation tasks. A starter-pack of modules is provided as an example of the conceptual design. This simple bias-reduction system may readily be applied in organisations and during everyday interpretations through to tasks for major commercial ventures.

**Keywords**

Cognitive bias, geoscientists, expert elicitation, conceptual design

**Significance Statement**

Experts are always required for the interpretation of geoscientific data. Data interpretation occurs on multiple levels and across a wide variety of tasks, from forecasting future climate given past conditions, or of future eruptions given observations of volcanic unrest, to the estimation of subsurface reservoir or solid rock volumes for a variety of purposes. The decisions made based on such interpretations often have a significant effect on society: from political policy or life-saving evacuation decisions, to deciding on commercially viable drill-sites.

Ideally decisions are made based on information about a range of best estimates and uncertainties of key quantities under different scenarios. Unfortunately, uncertainties in expert judgements are generally unknown. Different experts have different experience and access to different information, and cognitive biases in each expert results in sub-optimal interpretations, so interpretations vary from expert to expert.

The wide variety in geoscientific interpretation tasks and of experts performing these tasks led to the design of a Modular Interface for Cognitive bias in Experts (MICE). This highly flexible design has the potential to be used out-of-the-box with the starter-pack of modules provided here, with substantial room for expansion, and tailoring as required to a specific task or organisation with minimal effort. This work presents a simple bias-reduction system that can be readily applied in organisations trying to reduce cognitive bias in experts, ultimately to improve data interpretations and reduce uncertainty and risk in subsequent decisions.

## Introduction

Human judgement is an essential asset that is used to tackle complex geoscientific tasks for which computational techniques are insufficient (e.g., Baddeley et al., 2004; Kastens et al., 2016). Examples include the geological interpretation of a seismic image of the subsurface of the Earth (Bond et al., 2012; Alcalde et al. 2019), the decision of where, or if, to drill a well, assessment of the potential for carbon capture and storage in a particular subsurface reservoir (Larkin et al., 2019; Michie et al. 2021), hazard and risk assessments in volcanology (Donovan et al., 2012; Whitehead & Bebbington, 2021), and in making climate change forecasts (Anderegg et al., 2010; Dessai et al., 2018). There are, however, several limitations to the use of human judgements. Humans all have different experiences so that individual humans make judgements based in different, limited information. Humans also have limited perception, memory, and processing power (Stibel et al., 2009; Spellman et al., 2015; Davis & Marcus, 2020; Meyerhoff et al., 2021), and are susceptible to additional factors such as emotion (Lerner & Keltner, 2000), ego (Pulford, 1996; Muthukrishna et al., 2018), and motivational drivers (Zuckerman, 1979; Halamish & Stern, 2021).

This results in some cues within data or a situation going unobserved (e.g., Wilson et al., 2000), failure to recall some possibly related knowledge or experience during a task (e.g., Tverskey & Kahneman, 1973; Arnold et al., 2000), considering or assessing only a subset of the range of potential solutions (Baron et al., 1988; Glaser et al., 2005), and the avoidance of more complex interactions when simple ones will apparently suffice (Erber & Erber, 2000; Isen, 2003). Thus, while humans can produce judgements from minimal data, and make decisions quickly, this advantage is offset by the fact that these decisions may be suboptimal, and judgements may not best reflect even an individual expert's knowledge and experience. This suboptimality prevails in both individual and group judgements (e.g., Einhorn et al., 1977; Kerr & Tindale, 2004; Polson & Curtis, 2015) and is commonly attributed to a set of cognitive biases (e.g., Kahneman & Tversky, 1996; Haselton et al., 2015).

Motivated by the desire to improve expert performance during data interpretation, this work presents the design of a method or tool intended to illuminate, and potentially reduce, any cognitive biases in play at the time of expert judgement. This tool is the Modular Interface for Cognitive bias in Experts (MICE) and is intended for broad-use within the geosciences where high risk, high gain problems are frequent, and where the questions being asked cannot be answered straightforwardly from available data, thus requiring expert interpretation and judgement to bridge this gap.

## Evidence of expert bias in the geosciences

Across the geosciences, variations in expert judgement have been noted for many decades (Cooke, 1991; Aspinall, 2010). However, it has only been in the last two decades that some of these variations have been formally linked to cognitive processes, rather than attributed purely to variations in expert knowledge (e.g., Rankey & Mitchell, 2003; Bond et al., 2007). Three distinct examples are presented below to illustrate the presence of cognitive biases inherent within the geosciences. In the following, we refer predominantly to the heuristics and biases classification of expert sub-optimality that stem from Tversky and Kahneman's seminal works (e.g., Kahneman & Tversky, 1973, 1977; Tversky & Kahneman, 1973, 1974, 1981, 1986). Namely – heuristics as subconscious rules-of-thumb, and biases when these heuristics lead to systematic errors (Kahneman et al., 1982). Definitions of some of the principal heuristics and biases referred to in the text are:

***Anchoring and Adjustment:*** The formation of an initial opinion or estimate (an anchor), that is subsequently updated as further information is considered (Tversky & Kahneman, 1974).

***Availability:*** The use of the most easily recallable frequency of events within memory as a proxy for the actual frequency of events ever encountered (e.g., Tversky & Kahneman, 1973).

***Confirmation:*** The tendency only to look for, or see, further information that supports a current theory (e.g., Nickerson, 1998).

***Overconfidence:*** The tendency to place more confidence than is justified in personal judgement, thus causing the narrowing of uncertainty ranges or the allocation of higher probabilities to estimates (e.g., Glaser et al., 2005).

***Representation:*** The comparison of observed cues to known objects or situations to determine which option is best represented by these cues (e.g., Tversky & Kahneman, 1974; Kahneman & Frederick, 2002).

However, we note that the findings and problem definition do not depend on the adoption of this classification scheme, nor does the subsequent solution design.

### *Example 1: Reservoir Predictions*

Rankey & Mitchell (2003) set out to quantify the impact on reservoir predictions (for oil and gas exploration) due to inter-expert variation during seismic interpretation. Real seismic data from a Devonian pinnacle reef in the Western Canadian Basin were presented to six experts. The base horizon (bottom) of the reservoir was fixed, and the experts were asked to interpret the location of the top of the reservoir interval based on the seismic data, and two well ties. Once an initial interpretation was made, the experts were provided with further data. This included information on the geology of the reservoir, and six additional well ties. The experts were then asked to perform their interpretation again, given this new information.

Experts indicated that the task was relatively straightforward, but the results from both initial and final estimates showed large inter-expert variability in some regions of the seismic section. Additionally, despite the large amount of additional information provided for the second estimate, experts made minimal changes from their initial interpretations, with two experts making no changes at all. Rankey & Mitchell (2003) suggested that seismic interpretations are based on both tangible (data type and quality) and intangible elements (expert experience, preconceived notions, and geological understanding). The results suggest an *Anchoring and Adjustment* heuristic related bias (Tversky & Kahneman, 1974), with experts over-anchoring to their initial estimate (*Confirmation bias* – Nickerson, 1998).

*Example 2: Recharge Models*

A nuclear waste site at Yucca Mountain (Nevada, U.S.A.) was proposed that required contaminant transport (via groundwater) predictions to determine any detrimental effects to the surrounding areas. Ye et al. (2008) were tasked to determine which of five hydrological models were most realistic for this problem, for which they relied upon seven experts. These experts were selected based on their familiarity with the specific region's hydrogeologic conditions, and their proven research excellence within the specific conditions and location. Five of the seven experts were classified as specialists (each with expertise in a specific technical aspect of the problem), and two as generalists (experts in hydrology with general understanding of technical aspects).

A month before the elicitation of experts' judgements, information was sent out on the five models and the region of interest, as well as more general information on expert elicitation procedures, probabilities, and model uncertainty. During the month, some experts requested further reading materials or discussions on individual models for clarification, all of which were provided by the elicitation facilitators (Ye et al., 2008). Immediately prior to the elicitation, further training was given on the problem, as well as information on three cognitive biases (*Overconfidence*, *Anchoring and Adjustment*, and *Availability*). The experts were also encouraged to interact when discussing the models and any of the prior information.

Individual judgments were obtained via an elicitation of seventeen questions that were both qualitative and quantitative. Despite the abundance of information and inter-expert discussions, opinions about the plausibility of individual models were distinctly different between experts. This work by Ye et al. (2008) illustrates the danger of using a single expert, but also highlights an issue in the use of multiple experts: this was a well-designed and implemented elicitation of world-leading experts on the problem, and yet major discrepancies remained between expert judgements. Ye et al. (2008) did not note any *Overconfidence* bias during this task but could not discern whether the other two (*Anchoring and Adjustment*, or *Availability*) biases had any influence on the result.

*Example 3: Seismic Interpretation*

During most seismic interpretation, the true answer is rarely ever known; rather, it is inferred from many drill sites, or numerous seismic surveys (e.g., Brown, 1986; Macrae et al., 2016). To avoid an unknown answer complicating their uncertainty investigations, Bond et al. (2007) created a 2D geological model based on an initial layer-cake stratigraphy and subjected it to an inverted growth fault through forward modelling based on a set of geological assumptions. From the final cross-section, a synthetic seismic section was generated and given to 412 experts who were each asked to provide a single interpretation.

The experts had a range of experience levels (from master's student level to 15+ years) and a range of specific seismic interpretation backgrounds (e.g., salt environments, thrust-faulting, extensional settings). Of the experts, only 21 % identified the correct tectonic setting, and only 23 % identified the three main fault strands. Evidence for expert bias (referred to as "Conceptual uncertainty" by the authors) was identified by positive statistical correlations between a participant's answer and their area of expertise. This relationship was found in all experts except those with a strike-slip expertise (Bond et al., 2007). For example, a PhD student studying *salt tectonics* identified a *salt body* within the seismic section, an expert with 15+ years' experience in *thrust tectonics* observed only a *thrust fault*, and a similarly experienced *extensional expert* observed only *extension*.

The difficulties encountered by the experts in the experiment of Bond et al. (2007) can be linked to two heuristics: *Representation* and *Availability* (Tversky & Kahneman, 1973, 1974). During an interpretation, an expert is simultaneously looking for shapes or patterns (cues) and comparing these to known structures and possible overarching interpretations accessible within their memory. Which of these memories are accessed at any instant in time falls under the *Availability heuristic*, and the perception of cues and how well these reflect known structures is the *Representation heuristic*. Under conditions of high uncertainty (as for the task set up by Bond et al., 2007), experts may rely more heavily on their existing or available recollection of structures causing an *Availability bias* and may misperceive cues within the seismic section as reflections of these better-known interpretations (*Representation bias*).

Unfortunately, few if any conclusions can be drawn from this work in terms of how experts approach interpretations in the real-world. The elicitation expressly asked only for a single interpretation, without any corresponding confidence (uncertainty) level, and without any geological background information or additional data sources that usually accompany such interpretations. In subsequent work, the authors investigated experts' interpretation processes and found that those experts who considered how their interpreted structures could have evolved into existence over geological time (4-dimensional interpretation) were more likely to be successful (Macrae et al., 2016). They also found that once a conceptual model was envisioned, there was no evidence of deviation or reassessment of this model given further conflicting data (Alcalde et al., 2019).

### *Summary*

The above examples provide evidence for *Availability* and *Representation* biases (Bond et al., 2007), and *Over-anchoring* bias (Rankey & Mitchell, 2003) within geoscientific interpretations. Also highlighted were the potential dangers in the use of a single expert, irrespective of their level of expertise (Ye et al., 2008). The aim of this work is to provide a way to improve the performance of available experts, thus, the logical next step is to investigate where and how these biases may enter an expert's judgement and interpretation process.

While observations of cognitive bias within these highly controlled environments are common, exactly how to observe the presence of any potential biases in real-life applications remains elusive. This is further complicated when a project is spread over both time and space, and involves multiple personnel performing various roles – as is commonly the set-up for many geoscientific ventures.

Furthermore, the identification of the presence of a certain, or set of, cognitive biases may not be sufficient, as awareness of bias does not always reduce its influence (Pronin et al., 2002; Ehrlinger et al., 2005). Thus, quantification of the influence of various cognitive biases is also highly sought after, as this may then be backpropagated through the system to aid the estimation of true uncertainty in a judgement.

If a solution could be constructed to the problems induced by cognitive bias in geoscientific settings, then risk and uncertainty estimates that depend on expert judgements could be more heavily relied upon. Experts would be more likely to perform more consistently and at an improved level, which in turn will lead to better outcomes and lower risk.

## Design definition

To build a solution, it was necessary to first identify the current state of knowledge within which the problem, and subsequent solution lay. This identification can be used to distribute the focus of initial work, for example, between the time spent on in-depth investigations into advantages and limitations of existing solution attempts, and research into theoretical advances that may be applicable. The Knowledge-Innovation Matrix (Gregor & Hevner, 2013, 2014; Fig. 1) can be used for this and is based on the extent of knowledge about both the problem and potential solutions. Existing knowledge for this work includes research predominantly in the psychological literature on cognitive biases (e.g., Tversky & Kahneman 1973, 1974, 1981; Hogarth, 1975; O'Hagan et al., 2006), as well as guidelines on expert elicitation and information gathering techniques (e.g., Linstone et al., 1975; Wood & Ford, 1993; Cooke & Goossens, 2004; Knol et al., 2010; Oakley & O'Hagan, 2010; Truong et al., 2013). However, a large amount of application specific knowledge remains unknown.

In the geosciences, although there is strong evidence for the existence of cognitive biases during expert interpretation (e.g., Rankey & Mitchell, 2003; Bond et al., 2007), quantification of these biases and specifics around when and how they enter an interpretation process remain poorly defined (e.g., Stewart & Lewis, 2017; Pérez-Díaz et al., 2020), thus problem knowledge is low (Fig. 1). This is not necessarily a bad thing as designs that stem from low problem knowledge tend to be more flexible and more widely applicable (e.g., Simon, 1996). However, solutions then need training or tailoring as the problem becomes more defined, or as a case-study is presented (e.g., Mascitelli, 2000).

Solution knowledge is also low (Fig. 1). If multiple experts are available then structured elicitation methods may illuminate variations between experts to estimate uncertainty in the results (e.g., Linstone et al., 1975; Aspinall & Cooke, 2013). Computational methods may be used to optimise next-question choice during an elicitation to reduce uncertainty but tend to ignore the problem of cognitive bias (Curtis & Wood, 2004; Runge et al., 2011). Polson & Curtis (2010, 2015) and Arnold & Curtis (2018) propose structured quantitative methods to monitor information flow between experts and to detect potential biases during group elicitations. These works showed that these more structured group methods lead to quite different results compared to those from elicitations conducted with the same set of experts individually, but so far these methods have undergone only limited testing.

The main outstanding issue with all the above potential solutions is that a formal elicitation structure is required for implementation. This is not only time-consuming, but also is rarely practical during most real-world geoscience interpretation problems (Baker, 1999; Curtis, 2012). A more flexible, and tangible solution is required. This design development then followed the first three steps of a Design Science Research Methodology (DSRM) process model (e.g., Peffers et al., 2007), specifically:

1. Identify Key-Problems

2. Define Solution Objectives / Design Principles

3. Design and Development

The first two are presented below; the third is given in the following section (Modular Interface for Cognitive bias in Experts).

*Identify Key-Problems*

The motivating problems for this work can be defined as follows:

*Key Problem 1: The use of expert judgement in geoscientific interpretation is necessary and valuable but may be biased.*

The presence of cognitive bias may lead to risk assessments that do not always capture the true uncertainties, and to decisions being made with suboptimal information. This is a key-problem across many geoscientific disciplines and results in wells being drilled that produce outside of their predicted production ranges (Uman et al., 1979; Lerche, 2012), unexpected tornadoes (Simmons & Sutter, 2005; Jolliffe & Stephenson, 2012), and large earthquakes occurring in locations assigned negligible risk (Stein et al., 2012) or with ground-motions well above model predictions (Reyners, 2011). The level of cognitive bias varies with time, expert, and any external interactions (e.g., with other experts) that may lead to further group biases, thus the influence of bias on an expert's judgement can be viewed as continuously varying (Polson & Curtis, 2010), and the variability may be substantially enhanced by a lack of staff continuity (relatively common over longer geoscientific projects).

*Key Problem 2: Experts do not self-correct for cognitive bias.*

Cognitive biases stem from the subconscious, rendering self-awareness of their presence and influence infeasible (e.g., Duval & Silvia, 2002; Winne & Azevedo, 2014). Even with external prompts (e.g., from the elicitor or another expert), awareness and acceptance of individual cognitive biases within their own judgements is low amongst experts (e.g., Pronin et al., 2002; Ehrlinger et al., 2005). Thus, while experts may hone their expertise with time, and expand or enhance their knowledge base through experience, the level and influence of cognitive bias does not correspondingly reduce, and under some conditions, may even increase due to a stronger reliance on their prior knowledge (e.g., Bond et al., 2007).

*Key Problem 3: The use of experts is likely to increase in many geoscientific sectors.*

The use of experts, and therefore the influence of cognitive bias on a process, is likely to increase in the near future in the geosciences. Examples include: (i) the energy industry, as new ventures are pushed into unknown territories accompanied by larger uncertainties (e.g., deep offshore drilling – Zhang et al., 2019; or carbon capture and storage – Mikulčić et al., 2019), and (ii) climate change forecasts, as mitigation and prevention measures are introduced, their influence on future local and global climates need to be estimated far into the future where data is lacking (e.g., al Irsyad et al., 2019; van Vuuren et al., 2020). Thus, without external intervention, rather than diminishing with time, the influence of expert cognitive bias on societally important and novel geoscientific ventures is likely to increase.

*Key Problem 4: How to observe, quantify, or reduce the influence of cognitive bias under real-world conditions is unknown.*

Most cognitive bias observations come from lab-based psychological studies (e.g., Holleman et al., 2020), and the evidence of cognitive bias within the geosciences comes from highly structured, and well-controlled elicitations (Rankey & Mitchell, 2003; Ye et al., 2008; Polson & Curtis, 2010). Within real-world research and commercial environments, the observations (and observation techniques) for cognitive bias are lacking. These environments may be vastly different to lab-conditions, with high numbers of varying personnel, complex interactions (e.g., from formal emails to watercooler discussions), and with projects or interpretation processes spread over both time and space (geographic locations of personnel). Thus, the observation problem is to identify highly detailed characteristics (cognitive biases) within individual components (experts) that are continuously varying across space-time. The subsequent steps of quantification of the extent of this bias, and then the successful reduction or removal of effects, are equally elusive.

*Define Solution Objectives / Design Principles*

For the construction of a practical tool to tackle cognitive bias, eight driving design principles were identified:

*Design Principle 1: Flexible construction*

Tool construction should be able to both adapt to changing conditions and maximise future usage. A flexible design should allow for wider functionality and should accommodate both foreseeable, and unforeseeable future applications.

*Design Principle 2: Outputs should be independent of specific users*

The data collected or produced during any monitoring or analysis should not be biased by, or be in any way dependent on, the tool operator. No subjectivity should be introduced via tool usage.

*Design Principle 3: Intuitive operation*

Use of the tool should not require expert knowledge or extensive training.

*Design Principle 4: Draw attention only when necessary*

During any monitoring tasks, the tool should be minimally intrusive (both visually and audibly) to avoid any inadvertent influence on human behaviour. To minimise annoyance and/or time wasting, any other tool processes should only draw attention when necessary.

*Design Principle 5: Minimal processing times*

Tool run-time should be minimised wherever feasible. Any monitoring tasks must perform comfortably in real-time, and other tool processing times should match a project's pace.

*Design Principle 6: Negligible ongoing costs*

At project completion, the tool should have negligible ongoing costs, these include any hardware or software maintenance but also implementation and personnel training costs.

*Design Principle 7: Provide user choice and appeal*

Different user's seniority (e.g., data access) and preferences should be accommodated, and the tool should be visually and tactually appealing to encourage use.

*Design Principle 8: Meets all existing organisation regulations*

The tool should be compatible with all cybersecurity and commercial regulations, and confidentiality or personnel/employee agreements for data usage that may apply.

## Modular Interface for Cognitive bias in Experts (MICE)

Despite the large number of design principles, the position in the knowledge space (with a predominantly unknown problem and unknown solution) meant that the number of potential solutions remained vast. This was coupled with the fact that a single optimisation or selection criterion was lacking, and the absence of available observational information to indicate which solutions may be better than others, or even which solutions may work at all. To avoid a somewhat random tool creation with minimal justification, a modular architecture was designed whereby separate distinct components can be developed and inserted to suit the needs of the user, as well as allowing a high degree of design flexibility and the possibility for component replacement without significant overhaul of the entire system.

This modular architecture (Fig. 2) comprises four distinct module types:

**Monitoring modules** for observing and recording human behaviour (Table 1)

**Output modules** for storage and visualisation of observations or analyses (Table 2)

**Feedback modules** for analysis and comparison of observations (Table 3)

**Action modules** for direct intervention during a process (Tables 4a & b)

To create a coherent subset of modules for a specific application, **feedback** module requirements must match any **monitoring** module outputs, and the **output** modules must be compatible with both. **Action** modules can either be initiated by a facilitator, an expert, or in a computationally implemented set up - suggested automatically based on the results of **feedback** and **monitoring** modules. The tables presented for each module type are provided as an example, or starting set, for method application, and as such do not represent an exhaustive list of options for each.

*Monitoring Modules*

Observation methods of human behaviour are relatively well-defined (Phellas et al., 2011; Ciesielska et al., 2018) and range from completely passive methods (e.g., with the observers situated behind one-way mirrors - Sehgal et al., 2014) through to highly structured, specific tasks required of the participants (e.g., Aspinall & Cooke, 2013; Strickland et al., 2020). The methods suitable for monitoring modules in most geoscientific domains are constrained by practicalities, costs, and in most commercial cases - confidentiality restrictions. Monitoring modules can be split into two types – passive and active monitoring. While any observation is likely to have some influence on human behaviour (Alvero & Austin, 2004; Monahan & Fisher, 2010), this influence depends on monitoring method and tends to diminish with time as participants become accustomed to any monitoring equipment or personnel (Emerson et al., 2001).

Passive observation methods include video- or audio-recordings, and records of speech via computer or by hand (e.g., meeting transcripts or minutes). The completeness of the observation is an approximation for the amount of information that can be obtained from a passive monitoring method. For example, intonations, physical gestures, and the behaviour of participants who are not talking are all pieces of information that are lost during the use of transcripts alone. However, high-levels of passive observations, such as individual video-recordings are not only impractical and costly but can also be seen as highly intrusive and may result in lower levels of willingness to participate (Bottorff, 1994; Liu et al., 2015). The main limitation with purely passive observation methods is that the observer has no control over what information may be obtained. Thus, if a specific piece of data is required for each expert (e.g., "How likely do you think it is that this volcano will erupt tomorrow?"), passive methods cannot be relied upon. Another issue is the potential for observer error and/or bias. Information from passive observations must be acquired in real-time (unless video or audio equipment is permitted); thus, there is often only one opportunity to record the information, with no possibility to double check any observations. Depending on the number of experts, not everything said may be captured on a transcript, or in the minutes, and this problem is exaggerated in multi-lingual groups (common in commercial geoscientific ventures). Observer bias may be introduced during the selection of what (and what not) to record, and the presence of the observer may influence the behaviour of group members (Monahan & Fisher, 2010).

Active observation methods require facilitator interaction with the expert, or group of experts. These methods are usually applied in cases where specific information is required to be observed from the expert and are commonly formal expert elicitation methods (Cooke, 1991; Oakley & O'Hagan, 2010; Truong et al., 2013). A questionnaire is a widely applied elicitation method and may be used to gather information for a large variety of purposes from infrastructure risk assessments (Cooke & Goossens, 2004) to future volcanic activity (Donovan et al., 2012; Wadge & Aspinall, 2014). Although often more intrusive and time-consuming than purely passive observational methods, questionnaires ensure that specific information is obtained, and the results can be assessed multiple times by different observers, reducing the potential for facilitator bias. The main limitation of questionnaire use (aside from time) is that they must be carefully constructed to ensure all the required information is requested. How a question is phrased may influence an expert's answer (the *Framing effect* – Tversky & Kahneman, 1981; 1986), and the facilitator will likely require an expert's input into questionnaire construction. Implementation time is dependent on the number of questions asked, the complexity of questions, and the frequency with which the questionnaire is applied. In this modular design, a questionnaire design will likely be guided by any **feedback** module requirements, and it is envisioned that during application for some of these modules, a library of ten or twenty questions could be produced for an organisation or team, from which the user may select a project-appropriate subset.

An interview (rather than a stand-alone questionnaire) may provide more context to an expert's answers (Phellas et al., 2011), whilst still ensuring the required information is obtained. However, interviews are significantly more time-consuming, require well-trained facilitators, and may introduce facilitator bias (Hildum & Brown, 1956; Beveridge, 2021). Inconsistencies in data collection may also occur when multiple facilitators are running interviews in parallel (Bailar et al., 1997). Here, we present five monitoring method examples (Table 1), with two presented as an example of how these general concepts may be converted into specific actionable processes (Figure 3).

*Table 1: Monitoring module examples*

*Facilitator: An independent person to observe, collect, and/or analyse information from the participants (experts); Expert: Task- or domain-specific expert currently undertaking the task.*

| MODULE | DESCRIPTION | REQUIREMENTS |
|---|---|---|
| Video recording | A video recording is taken of a meeting to allow both behavioural and speech cues to be observed. | Facilitator, computer & video-capture device. Potentially speech recognition software |
| Audio recording | An audio recording is taken of a meeting to examine speech cues. | Facilitator, computer & audio-capture device. Potentially speech recognition software |
| Meeting transcripts / minutes | Complete transcript of meeting taken during the meeting, or minutes taken during, and confirmed after. | Facilitator/designated expert. |
| Questionnaire | One or more questions are presented to an expert to obtain specific observations to complete on their own | Facilitator, time. Potentially: expert input to question design |
| Interview | One or more questions presented to an expert to obtain specific observations to complete during an interview | Facilitator, time. Potentially: expert input to question design |

***Output Modules***

A potentially infinite number of possibilities are available for output modules. However, the majority of these are relatively straightforward to grasp as they are essentially just different ways of storing, grouping, and communicating the results from the other module types. The four presented here were selected due to common usage and therefore more likely to be correctly interpreted. Overly complicated output modules should be avoided unless these are prevalent within the specific discipline and well-understood by the experts performing the task. The four here are 'spreadsheet', 'line graph', 'point-value', and '3D graphic' (Table 2), the latter of which may provide more insight into more complex data but is unlikely to be used in most practical applications for presenting expert variations. The set-up time for the first three is negligible as they can be pre-coded to respond to data in certain formats, and may be produced in existing, well-known software requiring minimal training. For example, setting up a Microsoft Excel spreadsheet to produce a graph, or to calculate an average. The determination of which output module is optimum for a specific dataset is less straightforward, for example, when is a 3D graphic more confusing than informative? Or when is a line graph misleading? It is also likely that different experts will prefer to visualise data in different ways, thus the suggestion is here to use a minimum of two different output modules to communicate any data to decrease the potential for misinterpretation. Figure 4 provides two examples of how these output modules could be presented within the modular framework.

*Table 2: Output module examples*

*Facilitator: An independent person to observe, collect, and/or analyse information from the participants (experts); Expert: Task- or domain- specific expert currently undertaking the task.*

| MODULE | DESCRIPTION | REQUIREMENTS |
|---|---|---|
| **Spreadsheet** | Data presentation in tabular format. Columns usually represent variables (e.g., parameters), Rows usually represent samples (e.g., experts) | Expert or Facilitator, Computer |
| **Line-graph** | Line-graph of data. For example – variations in a parameter with time, or single expert variations. | Expert or Facilitator, Computer |
| **Point-value** | Point-values, either represented as a graph (e.g., expert estimates with uncertainties), or as standalone statements (e.g., "Expert A's estimate overlapped with Expert B's by 20 %") | Expert or Facilitator |
| **3D - graphic** | 3D graph or image illustrating changes in multiple parameters simultaneously. | Specifically trained Expert or Facilitator, Computer |

*Feedback Modules*

The **feedback** modules take observational data from the **monitoring** modules, analyse them against a specific set of criteria, and then feedback to the experts and/or the facilitator the results via **output** modules, from which a set of **actions** may be recommended. Four examples are presented here (Table 3), with how these may be linked to specific **monitoring** and **output** modules for application. Two of these are shown in Figure 5 as an example of how these general concepts may be converted into specific actionable processes.

*Table 3: Feedback module examples*

*Facilitator: An independent person to observe, collect, and/or analyse information from the participants (experts); Expert: Task- or domain- specific expert currently undertaking the task.*

| MODULE | DESCRIPTION | REQUIREMENTS |
|---|---|---|
| **Consensus vs Individual** | Compare individual assessments between experts, and between each expert and the group consensus. Feedback (usually anonymous) group variations. | Facilitator, specific information |
| **Uncertainty** | Track answer variation for individual experts and group. Assess abrupt changes, look for herding behaviour, provide continuous feedback around uncertainty range in answers. | Facilitator, specific information obtained over time (i.e., multiple observations required) |
| **Individual influence** | Feed apparent individual influence back, to allow expert comparison with task-specific expertise. Influence proxies include expert 'airtime', expert assessment of other's expertise. | Facilitator, specific information (perceived expertise), observations of speech/behaviour beneficial |
| **External consistency** | Compare expert knowledge with an external (ideally global) database. Feedback areas where knowledge is lacking, where expert knowledge greatly overlaps or does not. | Facilitator, specific information, access to external database of task-specific knowledge |

There are multiple ways to analyse individual or group observations before being fed back into a process. Some of these may reduce cognitive biases, improve knowledge consistency, or facilitate the assessment of true uncertainty within a final judgement. Four are described in detail here with information provided as to where similar steps have been implemented before, the level of participant and facilitator involvement required, and any links to other modules.

*CONSENSUS vs INDIVIDUAL*

This module is the comparison of individual answers between experts, and between each expert and the group consensus. The feedback is usually anonymous and should present these variations. Studies within both cognitive and geoscience literature have shown that a group consensus may not be superior or even equal to the judgement of the best individual within the group (Einhorn et al., 1977; Baddeley et al., 2004), although this may depend on how often the group members perform tasks together (Lark et al., 2015). Furthermore, the comparison of individuals' opinions with a consensus often reveals a better estimate of the true group uncertainty (Polson & Curtis, 2010), and valuable information may be contained within the rationale behind why experts form differing opinions that may be lost (Ericsson & Simon, 1980). By assessing the variation in individual and group judgements at a single instance, *Overconfidence* bias may be observed (Glaser et al., 2005; Speirs-Bridge et al., 2010), and via assessment over time, *Herding* bias may also be noted (Polson & Curtis, 2010). By providing feedback (often anonymously) to a group about the range of individual opinions you may catch misunderstandings (Polson & Curtis 2010), improve the group results (Hemming et al., 2018), and obtain a better estimate of the true-uncertainty inherent within the problem (Hanea et al., 2018).

    **Monitoring** module: To track individual opinions, specific information must be obtained at designated points in the process. This requires a structured elicitation method such as a questionnaire (which may comprise of only a single question), and requires both a facilitator to pose the question, and collect the results, and the willingness of the group members to participate.

**Output** module: Results from this feedback module can be straightforwardly presented on a line graph.

*UNCERTAINTY*

This module looks at answer and accompanying uncertainty variation over time for both individual experts and groups, variations should be assessed by looking for abrupt changes and herding behaviour, and relatively frequent feedback should be provided around these uncertainty ranges.

As an expert receives more data, they tend to become more certain about their judgement, this is often irrespective of how informative these data are (e.g., Nickerson, 1998). If a single variable, or parameter is continually assessed over the lifetime of a process, the uncertainty in this variable can be tracked, and then the evolution of this variable to-date fed back to the group at set intervals. This allows group members to re-assess whether their current estimate of uncertainty is logical, identify where and why abrupt changes in uncertainty were observed (e.g., due to new data, or a change in expert; Polson et al., 2012), and subsequently get a better idea of the true uncertainty. It may also aid the assessment of group or individual *Overconfidence* bias (Glaser et al., 2005).

**Monitoring** module: To provide this feedback, a semi-continuous assessment of a parameter of interest is required, as well as accompanying assessments of a perceived uncertainty range within which the parameter may lie. This module therefore requires a structured elicitation method, as well as a clear explanation to the experts around the definition of uncertainty, and how upper and lower (or percentile) ranges are defined within the context of the task. An alternative to a questionnaire that may better capture individual uncertainties is that of one-on-one interviews, however these are more intrusive to the main process and significantly more time-consuming.

**Output** module: Uncertainty ranges are commonly presented as line graphs with error bars;

however, point-values may also provide insight. For example, a statement that "the final parameter estimate lies within the 95$^{th}$ percentile of the initial uncertainty range".

*INDIVIDUAL INFLUENCE*

During group tasks, there is evidence to suggest that those people with the most problem-specific expertise do not necessarily have the greatest influence within a group (Baddeley, 2013). This is somewhat alleviated where a group is formed of people who are well known to each other (Lark et al., 2015). There is also evidence that experts with a narrower range of experiences are more likely to have stronger opinions (Kahneman & Klein, 2009), and consequently may assert greater influence during group tasks if not explicitly noted and externally adjusted for by a facilitator (Quigley et al., 2018). Feedback as to different individuals' influences on a group may provide valuable information as to whether individual influence is proportional to specific-task expertise. The information obtained through this module may also be linked to the CONSENSUS vs INDIVIDUAL module to moderate any observations of *Herding* bias (Baddeley et al., 2004) and to help calibrate the link between any observed behaviours and an individual's influence.

**Monitoring** module: In poorly controlled environments (i.e., real-life), any direct measurement of individual influence is difficult. However, other observable parameters may be used as proxies for influence. These might be the behaviour of a group when different members are absent, the amount of 'airtime' a person takes up, and the frequency of certain keywords or phrases amongst a person's speech (e.g., "I believe" vs "I agree"). It should be noted however that for these observations to be collected, the participants must be unaware of the specificities of the information being captured, to avoid any self-regulation of speech. Experts can also be asked to assess their own expertise relative to the others within the group. Thus, both questionnaire-based and passive monitoring methods may be coupled to this feedback module. The former of these requires active participant and facilitator involvement with the questionnaire likely to comprise a set of questions.

The latter requires no direct participant involvement but, depending on the proxies selected, recording equipment (and speech recognition software) may be required and/or the facilitator may need extra training and practice to reduce user-errors during data collection.

**Output** module: Line graphs may be used to illustrate individual's assessment of other expert's expertise and point-values may also be appropriate for proxy results (e.g., "Expert A spoke for 80 % of the group discussion time"). This is also a module for which a 3D graphic may be useful, especially if the results are observed to vary over time, or throughout a process.

*EXTERNAL CONSISTENCY*

This module is the comparison of expert knowledge within individuals and the group with an external (ideally global) database. Areas where the expert knowledge is lacking, and where experts overlap and do not are then fed back to the group. When performing a task under high levels of uncertainty, a set of guidelines or rules-of-thumb may be used and are often stated explicitly. For example, in a project looking at multiple potential drill sites, a guideline might state that "a reservoir is not producible in areas where porosity is below 10 %" – as such, any potential site with porosity lower than 10 % can be automatically ruled out, regardless of any other parameters (such as reservoir volume, depth, or temperature). These guidelines ensure that a moderate level of external consistency is imposed across projects or regions and help to reduce any expert variations around these explicitly stated parameters.

Experts also use subconscious rules-of-thumb that are likely expert-specific and based on individual experiences and training (Shanteau, 1992; Curtis, 2012; Chi et al., 2014). The information that they access may vary with their working memory, and therefore with time (Tversky & Kahneman, 1973), and may be influenced by external factors such as the order that data are presented (Murdock Jr, 1962; Arnold et al., 2000), other experts (Baddeley et al., 2004), and any motivational or emotional biases (Zuckerman, 1979; Halamish & Stern, 2021). Thus, an expert is providing a judgement based on their available memory of their specific knowledge base to tackle a task that ideally would be answered using all information contained within all (global) knowledge.

For example, an expert may be using their answer to the question "How often *do I remember* having seen this fault structure?" as a proxy for "How *common* is this fault structure?", the result from which is used to answer the task at hand: "How *likely is it* that this fault structure is here?"

If a comparison can be made between an expert's knowledge range of a parameter of interest and a global (or external) database, this may lead to a better estimate of the true uncertainty. It may also be possible to identify where a specific expert will be most valuable and areas where they may require additional training. Feedback may be provided before a process to identify individual expertise levels, after a process is complete to aid the assessment of uncertainty within a judgement, and during a process – for example during expert disagreements to aid identification of why their judgements may vary.

**Monitoring** module: Specific information is required for this feedback module and as such, passive monitoring methods are unlikely to be sufficient. Questionnaires or individual interview techniques are more suited to obtain a knowledge range from an individual, with roundtable or brainstorming sessions also valuable to ascertain group knowledge. An additional consideration for external consistency is that access is required to an existing database used to compare the knowledge of an individual or group against. This might be organisation specific (e.g., the productivity of existing well-sites in a region), or come from published research (e.g., a flood database containing all major floods in the last 100 years). Determination of which of these datasets are most applicable to the specific task will likely require some expert-facilitator interaction before module implementation.

**Output** module: Point-value results may be useful for this module for comparison of, and between, individual experts. For example, "expert A's estimate of how likely this fault is to occur is 10 %, global database says 20 %"; or "expert A and B's knowledge overlap for this data is 95 %".

More complex information may be better presented as line-graphs, or potentially 3D graphics but this will depend greatly on the specific module and data structure.

*Action Modules*

It cannot be assumed that the provision of feedback is sufficient for an expert to self- or group-correct any biases (e.g., Pronin et al., 2002; Ehrlinger et al., 2005). Thus, a set of action modules are the core of this modular design and are split here into *training* and *tool* modules that may facilitate the de-biasing of experts during a task. These action modules contain methods that may be undertaken regardless of whether any suboptimal expert performance is observed, and at least one training action module is recommended before any expert elicitation. Four examples are presented here of *training* action modules requiring varying degrees of participant commitment (Table 4a), and sixteen examples of *tool* action modules (Table 4b), three of which are described in detail below, and examples of these as actionable items are shown in Figure 6.

*Table 4a: Action module examples – Training*

*Facilitator: An independent person to observe, collect, and/or analyse information from the participants (experts); Expert: Task- or domain- specific expert currently undertaking the task.*

| MODULE | DESCRIPTION | REQUIREMENTS |
|---|---|---|
| **General bias awareness** | Informative online or in-person course on common cognitive biases, cues to look out for, and techniques to help reduce bias | Facilitator / Computer |
| **Job-specific bias awareness** | Informative online or in-person course on job- or role-specific biases, cues to look out for, and techniques to help reduce bias | Facilitator / Computer |
| **Tailored bias awareness** | Online or in-person cognitive tutoring course that constructs a learner's profile, and then provides specific bias information and reduction techniques | Facilitator / Computer |

| Simulation | Online or in-person training in which individuals / groups perform complex job-specific tasks that involve one or more cognitive error traps | Expert & Facilitator / Computer |
|---|---|---|

*PRE-MORTEM*

A pre-mortem approach to tackling the cognitive bias of *Overconfidence* (Glaser et al., 2005) was first suggested by Klein (2007) and draws on the knowledge of an observed phenomenon called the *Hindsight* bias. This bias is the overestimation of the likelihood of an event occurring, given the knowledge that it did occur (Christensen-Szalanski & Willham, 1991). Experts are asked first to make a judgement or plan of action, and then to imagine that this is found to be incorrect, or that the plan failed. They are then asked (individually) for reasons why this may have happened, and the results shared with the group. This process may widen the possible avenues for group or individual thought resulting in a potentially superior solution, as well as aiding a more accurate estimate of the true risk or uncertainty associated with a final judgement or proposed plan (Kahneman & Klein, 2009).

*DECONSTRUCT TASK*

Complex tasks are often presented to an expert, or group of experts, for which multiple types of data are available in varying amounts, that require ultimately a single answer or estimate. Task ambiguity may affect how a task is interpreted resulting in different experts essentially addressing different tasks (e.g., Einhorn & Hogarth, 1985; Ritow et al., 1990). It has been suggested that explicitly breaking a highly complex task into smaller, sub-tasks may improve the accuracy of an expert's final estimate (Hogarth, 1975). Assessment of task ambiguity may illuminate the differences in expert results, especially when the task is comprised of open-ended questions. Where and why expert judgements differ may be explained through answer justification ("why do you think this?"), either in the context of group discussions or during one-on-one interviews with a facilitator. As well as increasing the likelihood of capturing any expert errors from this justification information (Polson & Curtis, 2010), several underpinning subtasks or divisions may be apparent.

Asking experts to address these sub-tasks individually, and/or under various assumptions (e.g., "assuming that this horizon is a fault, what is your reservoir volume estimate?") provides more detailed, and often more valuable information than does application of a single, complex task (e.g., Rankey & Mitchell, 2003). This includes information as to where the largest uncertainties or inter-expert discrepancies lie, and may suggest the most beneficial areas for subsequent data collection. These sub-task answers can then be recombined (e.g., Arnold & Curtis, 2018), to provide more informative answers to the larger task, and a better estimate of the true uncertainties inherent in each of the experts' judgements.

This action module requires a facilitator for the main task who should have an in-depth knowledge of all task dimensions (i.e., be an expert themselves). Application of this module requires additional time as the group of experts are asked to approach the same task from two or more angles. However, the pay-off may be substantial in terms of reduction in expert misunderstandings, improved uncertainty estimates, and identification of task components that may require further data. This is amplified in cases where a decision is made under high levels of uncertainty and economic and/or societal risk.

*RISK ATTITUDE PROFILE*

A person's attitude to risk may influence both their decisions and judgments, especially of the likely occurrence of an event, or the profitability of a venture (e.g., Howard, 1988). Assessment of an expert's risk attitude is relatively common-place in financial sectors where any prevalent attitude amongst the experts may substantially influence financial gains or losses (e.g., within the stock market – Keller & Siegrist, 2006). In most situations, the desired expert attitude is risk-neutral (Willebrands et al., 2012), as a task's true estimate or answer given the available data, and expert-knowledge is independent of a person's risk attitude.

Risk attitudes are likely to vary with time, external influence, and recent experiences (Weber, 2010), thus assessment of an expert's risk attitude should be on-going. As an action module, assessment will be via a carefully designed questionnaire. Initial construction and fine-tuning are likely to take considerable time to ensure it is suitable within a specific organisation's culture, and that the questions closely follow the main tasks usually undertaken (Charness et al., 2020). Several versions may also be required to better align with different expert roles, and sectors within an organisation. However, once construction is complete, this tool becomes an 'off-the-shelf' action module. To build up an accurate risk-attitude profile, the assessment should be applied both independent of a task, and immediately before (or during) a specific task. The results may then either be viewed by the task facilitator only, who may then adjust individual's judgements accordingly, or be provided to the participants during a task to motivate behaviour adaptation, or the self-adjustment of judgements.

## Discussion

This work addresses the common problem of cognitive bias in geoscientists. The main design challenge was to produce a tool that could be directly applied to a wide-variety of scenarios, whilst working in a low-problem low-solution knowledge space. This ultimately led to a modular design, with a set of practical tools that can be used alone or combined as modules as the Modular Interface for Cognitive bias in Experts (MICE). This section describes how the design was developed around the key problems, and how MICE meets the design objectives. Current design limitations are then discussed, as well as directions for future development.

### *Addressing the Key Problems*

The key problems identified in the use of experts in geoscientific problems can be summarised as: (1) experts are necessary but biased, (2) they do not self-correct these biases, (3) the use of experts is increasing, and (4) optimal observation, quantification and reduction methods for cognitive bias remain unknown.

Key problems (1) and (3) meant that any design solution could not remove or avoid the use of experts, thus any automated or machine-learning options intended to replace or reduce expert usage were not considered. Key problems (2) and (4) meant that we could not be certain of any direct bias observations or estimates made, and even if we could and presented these to experts, there was no guarantee that this alone would reduce the influence of cognitive bias. There is a lot of research in this space that may be informative, or may reduce cognitive bias, but the problem is that none are proven as globally effective. As every task and every expert is different, so should the solution be; as such, the view was taken that most published research with evidence of effectively reducing cognitive bias can be considered a potential tool.

The focus then of the design was on the **action** modules – which represent the set of tools that may reduce or illuminate cognitive biases. Any input / presentation requirements for these tools then formed the basis for **monitoring** and **output** modules, and any further analyses required of these then formed the **feedback** modules.

*Meeting the Design Principles*

The design principles are summarised as follows: (1) flexible construction, (2) outputs independent of user, (3) intuitive operation, (4) non-intrusive to the process, (5) minimal processing time, (6) negligible ongoing costs, (7) provide user choice and appeal, and (8) meets organisation regulations.

The modular approach is flexible (1), provides user choice (7), and can be set up to meet most organisation regulations (8) via the exclusion or adaptation of any modules deemed not sufficiently secure, or the removal of those requiring an external facilitator (noting that an internal facilitator is always feasible with specific training). Any influence of user subjectivity to the outputs (2) is negligible for most modules, except those requiring some interpretation steps – usually within the monitoring modules (for example, where the facilitator is taking notes, or looking for behavioural clues).

Whether the output is non-intrusive to the process (4) is dependent on the modules selected. However, most action modules are intrusive, requiring the experts to stop the process, perform the action module, and then return to the task. During the design process it was decided that limiting the options to only passive methods for reducing cognitive bias would be debilitating to the overall approach. However, with recurrent use of the MICE approach, and as familiarity grows with the different tools, it is envisioned that these modules would become part of the task, rather than a separate add-on, at which point they would be intrinsic, rather than intrusive.

Once set up, there is minimal processing time (5) and negligible ongoing costs (6) with all the MICE module examples presented here, dependent on how the facilitator aspect is managed (e.g., an external contractor is much more costly than an internal expert acting as facilitator). There is a likely unavoidable increase in task-processing time (and therefore personnel cost) with the addition of any bias related training, or modules, however, the benefits in output accuracy are likely to outweigh these.

Intuitive operation (3), and user appeal (7) have yet to be assessed, as this cannot be directly measured without external user input.

### *Limitations and Avenues for Future Development*

Two main limitations are noted here – the common requirement for a facilitator to run the module, and the set-up times of some of the more involved modules. The use of a facilitator within many of the modules remains a potential implementation limitation. This is both in the potential for facilitator biases entering the process, and in terms of monetary cost – requiring the employment and training of an extra person who also needs to be present during most of the task. While some of this cost may be absorbed by training existing personnel, these people may then not be able to participate in the task itself – so an expert may need to be traded for a facilitator. The implementation of those modules with more potential benefit, generally requires more commitment of the experts and the organisation to the process, and a well-trained facilitator. As such, there is a potentially non-negligible, an unavoidable set-up cost in both time and effort.

There are two logical next steps for future development of this conceptual design – the practical implementation of the method within a set of case-studies and expanding the module sets. Case-studies would ideally be a varied set, that span both research and commercial institutions, and with experts performing tasks from different discipline sets within geoscience. As there is no limit to the number of modules that can be added or developed for this, the set of module examples can be greatly expanded. While the modules presented here may be applied across all geoscientific tasks, it is feasible that more discipline-focussed modules may also be valuable. For example, specific visual cues that might be used during seismic section interpretation may benefit from a tailored training action module around the issues of representation bias, or from a tool action module that illuminates how an expert is performing their interpretation (e.g., a 'say what you see' approach). Both avenues would be beneficial to explore in future work.

## Conclusions

Experts are required for the interpretation of data throughout the geosciences. Data interpretation occurs on multiple levels and across a wide variety of disciplines, from forecasts of future climate change, or of future eruptions from observed volcanic unrest, to the estimation of reservoir or rock volumes for a variety of purposes. However, expert answers are commonly suboptimal, stemming from experts' inherent subjectivity, adverse external effects or motivations, and cognitive biases. The reliance on experts within geoscience is likely to increase in future. Thus, without external intervention, the influence of expert suboptimal performance on societally important and novel geoscientific ventures is likely to increase.

This work presented three examples of the suboptimality of experts prevalent in the geosciences and formally defined the overarching problem and design principles that any solution should follow. Based on these, the Modular Interface for Cognitive bias in Experts (MICE) was designed that combines four module types: monitoring, output, feedback, and action tools. Modules can be added or removed as required. Action modules can be used as stand-alone tools that may reduce or illuminate cognitive biases, or can be suggested via observations (via monitoring modules), or analyses (via feedback modules). This simple bias-observation and reduction system can be readily applied to geoscientific tasks that require experts, and the general concepts in this design can be converted into specific actionable processes that may be tailored to individual organisations. Further development of the MICE design should begin with varied case-study applications and expansion of the module starter set.

## Declarations

***Ethics approval and consent to participate***

Not applicable

***Consent for publication***

Not applicable

***Competing interests***

Not applicable

***Availability of data and material***

Not applicable

***Funding***

We thank Total UK for sponsoring this research.

*Authors' contributions*

MW designed the modular interface with input from AC. Both authors contributed to discussions around the idea and the writing of the manuscript.

*Acknowledgements*

Not applicable

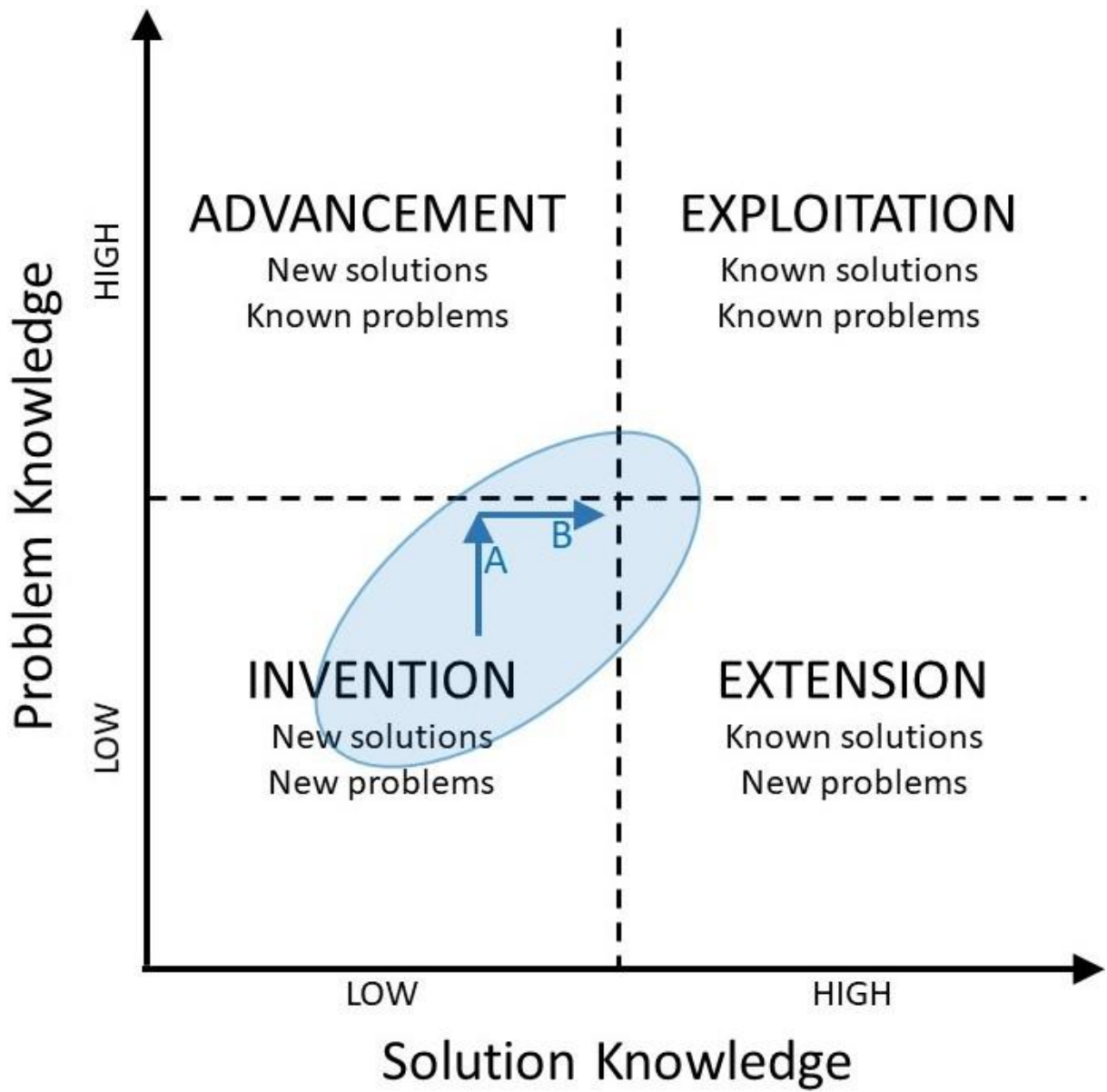

*Figure 1: Knowledge-Innovation matrix (adapted from Gregor & Hevner, 2015). The blue ellipse represents the relevant design area for this problem. The blue arrows represent potential increase in problem knowledge when applied to a specific problem or case-study (A), and the potential future increase in solution knowledge based on evaluation and feedback after initial tool deployment (B).*

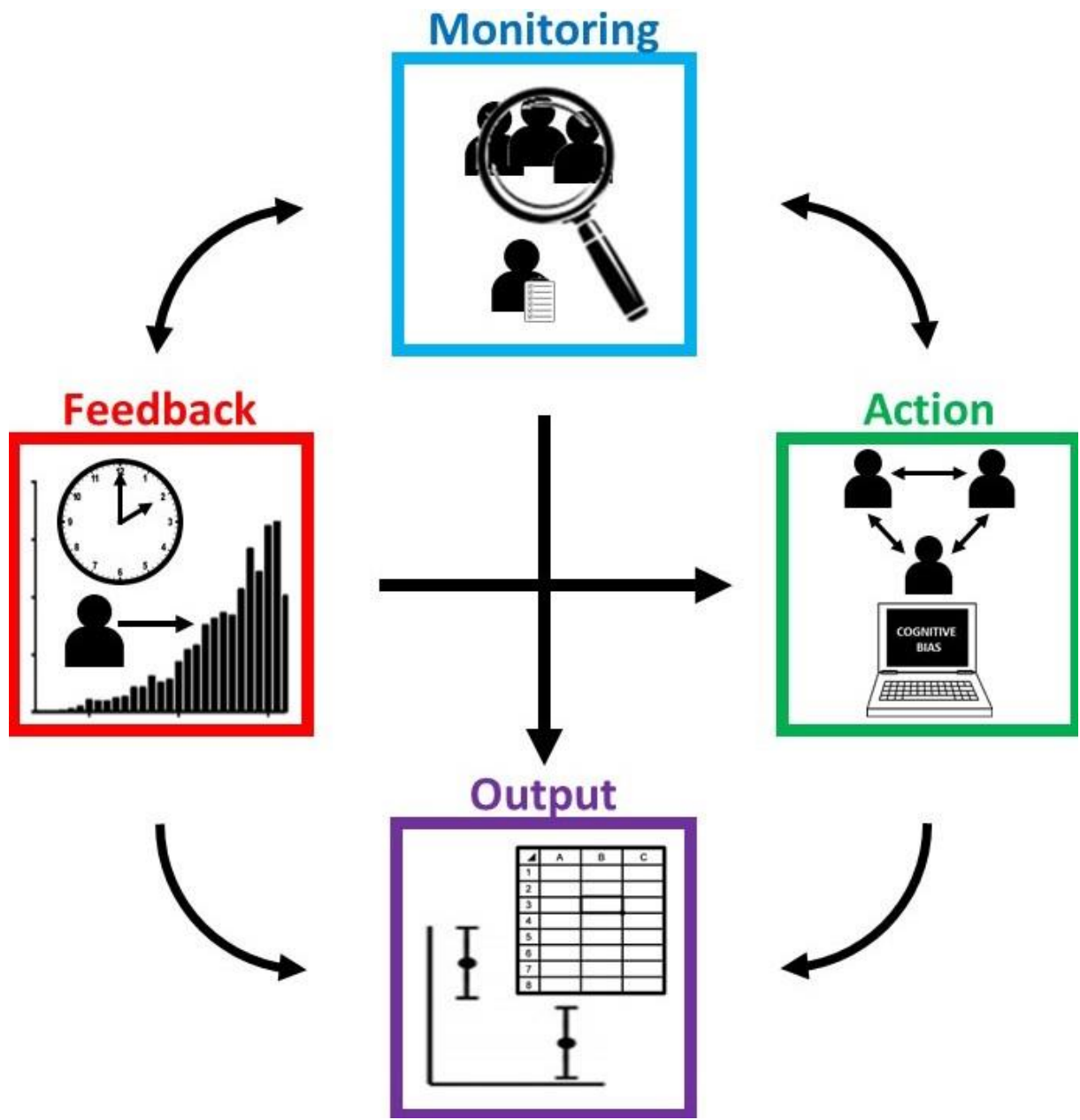

*Figure 2: Modular architecture of MICE (Modular Interface for Cognitive bias in Experts) illustrating the links between the four different modules: Monitoring, Feedback, Action, and Output.*

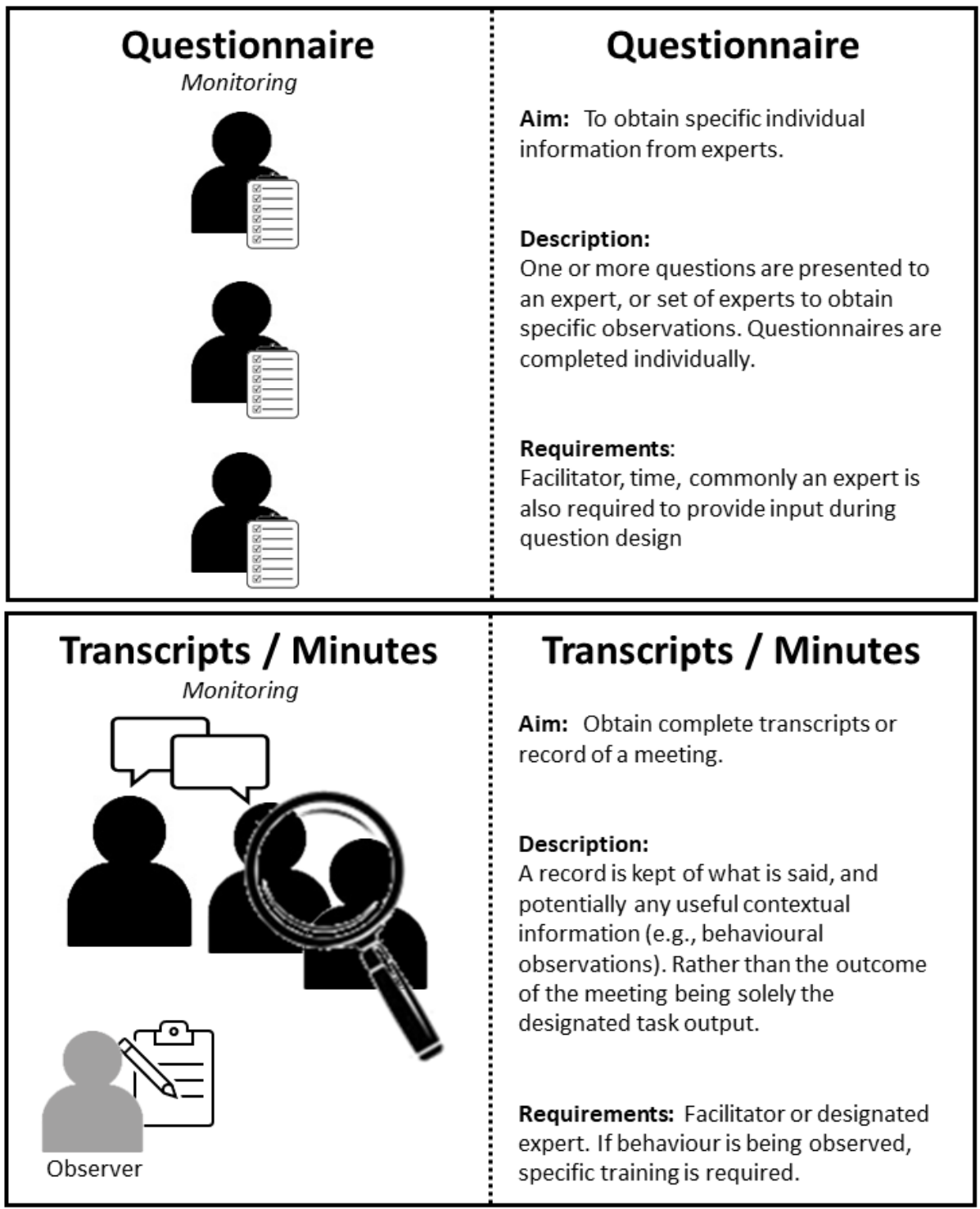

*Figure 3 – **Monitoring module examples.** Two of the conceptual monitoring modules are presented as actionable items (cards): the questionnaire, and the taking of transcripts/minutes.*

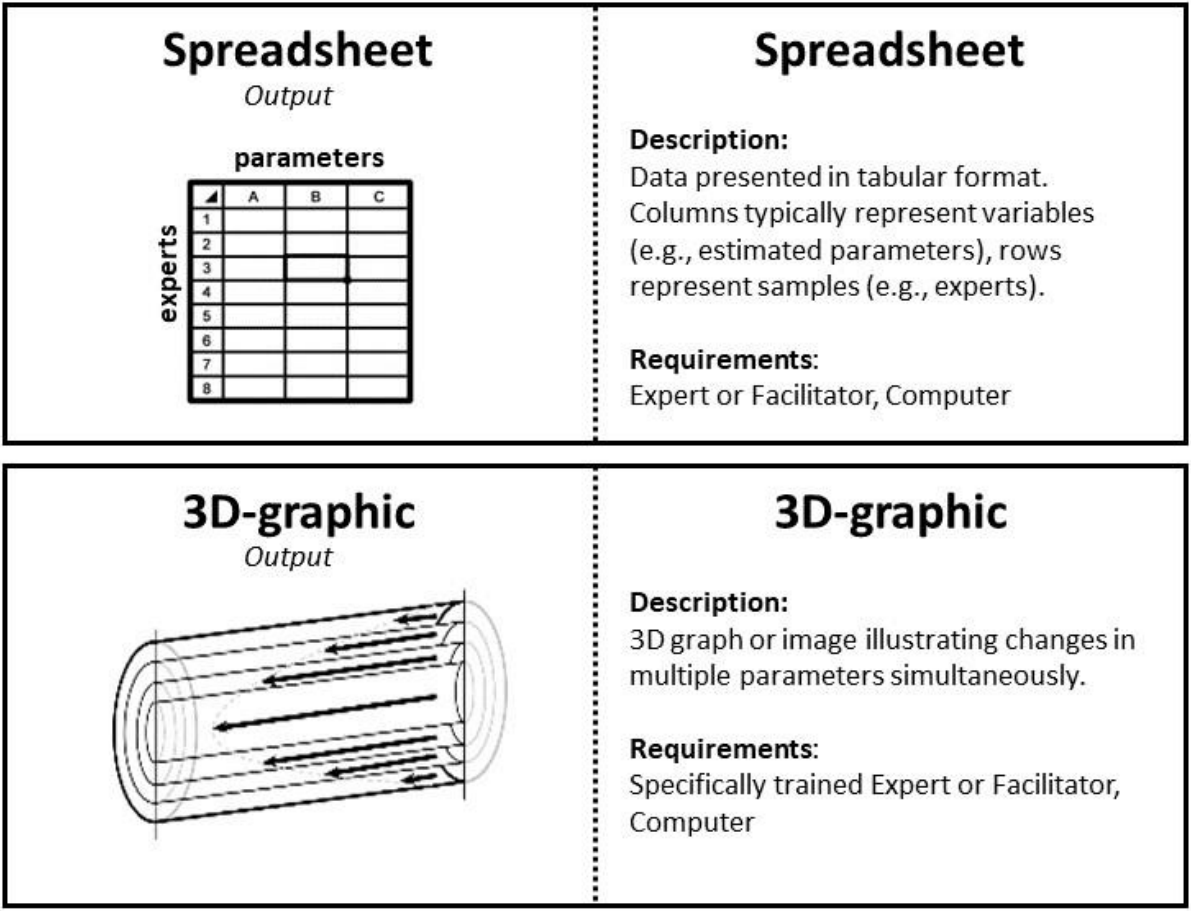

*Figure 4 – **Output module examples.** Two of the output modules are presented as visual aids: a spreadsheet, and a 3D-graphic.*

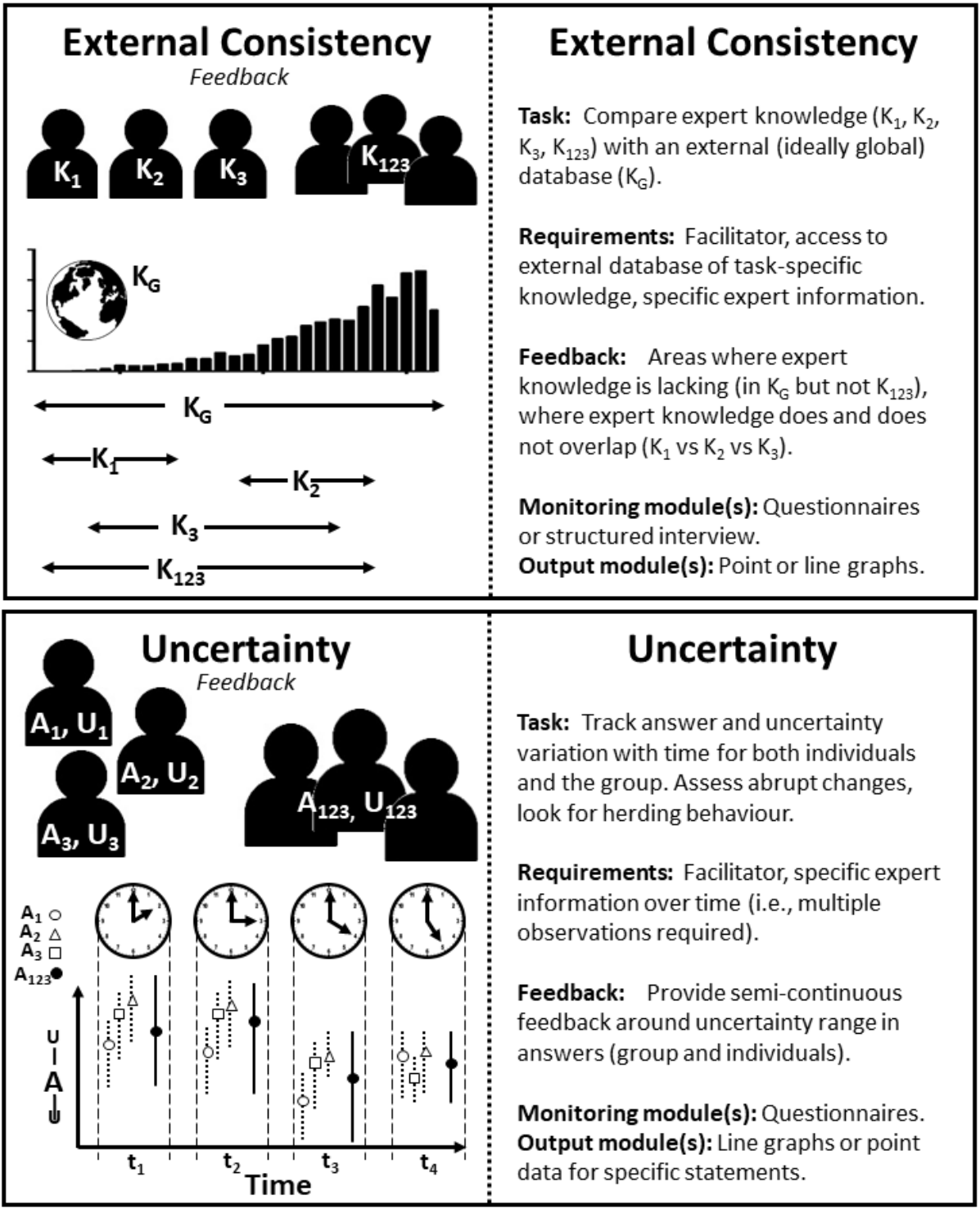

*Figure 5 – **Feedback module examples.** Two of the feedback modules are presented as visual aids: external consistency, and uncertainty. See text for detailed explanations of each.*

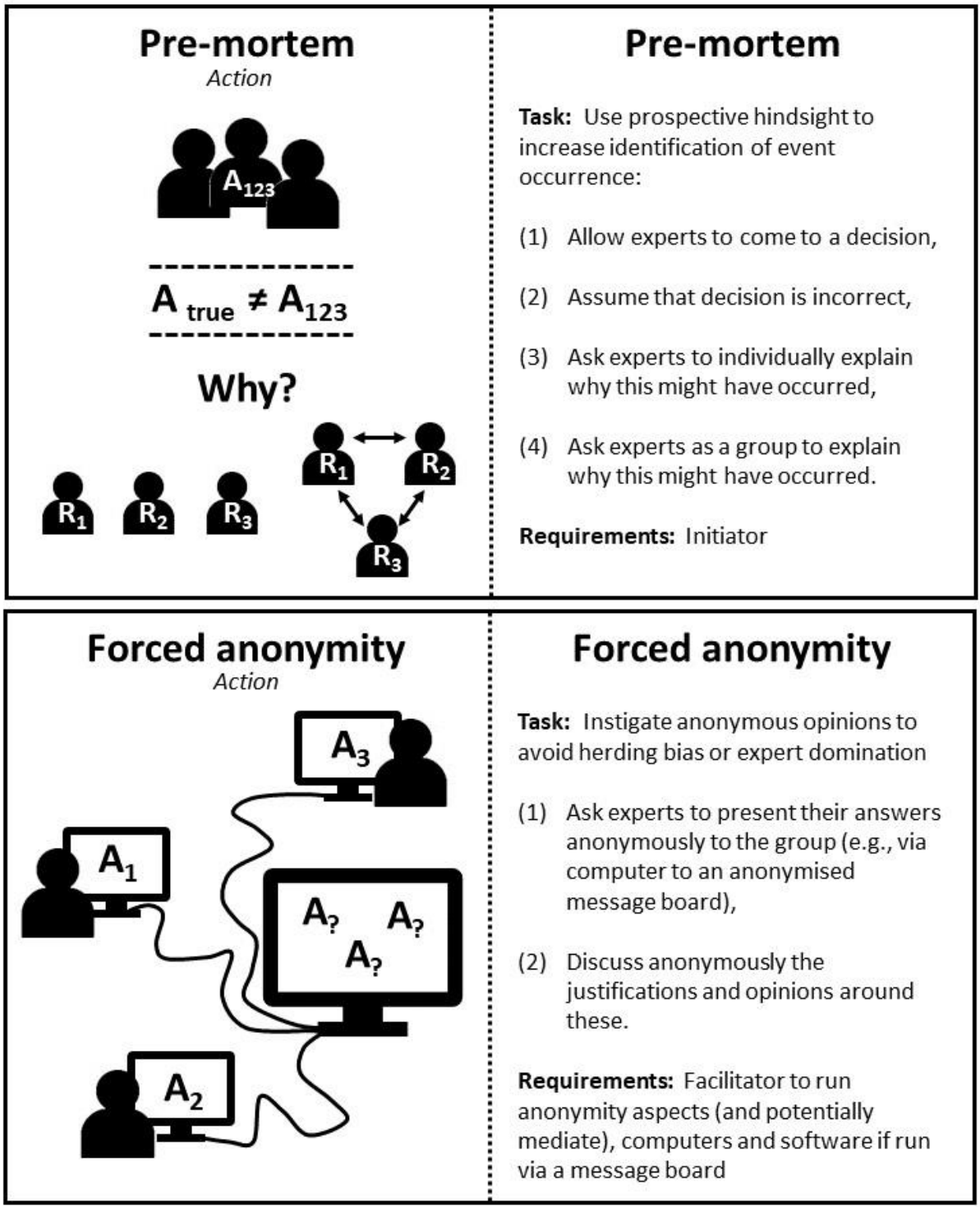

*Figure 6 – Action module examples.* Two of the action modules are presented: pre-mortem, and forced anonymity. See text for detailed explanations of each.

*[Table 4b: Action module examples – Tool]*

*Facilitator: An independent person to observe, collect, and/or analyse information from the participants (experts); Expert: Task- or domain- specific expert currently undertaking the task; Initiator: A participant (expert) who suggests the action; Problem-owner: Either the head of the project, or research leader who formed the task.*

| MODULE | DESCRIPTION | REQUIREMENTS |
|---|---|---|
| **Data check-list** | Structured data acquisition: Instigate process that forces deliberate data acquisition, and aids identification of missing data | Creation of, and access to data check-list |
| **Step-back** | To encourage metacognition: Ask experts to discuss why they have made a specific judgement or decision, including any relevant experiences or knowledge. | Initiator |
| **Slow-down** | Forced time-out: To allow time for reflection and to avoid early consensus or group herding. Ask experts to take 5-10 minutes in which the subject is not considered (e.g., tea break). | Initiator |
| **Ask again later** | Ask the same expert(s) to address the same problem / decision at a later time or date. Then ask if the expert thinks their answer is consistent with their previous answers. | Facilitator |
| **Pre-mortem** | Use prospective hindsight to increase identification of reasons for event occurrence by assuming a plan or judgement has proven to be incorrect (see text for further details) | Initiator |
| **Seek advice or knowledge** | Consult expert(s) outside of the current working group for information in identified regions of high uncertainty or lack of specific expertise within the group | Access to external experts / database |
| **Devil's advocate** | In team-based decisions, designate a member as devil's advocate to force additional options / justifications by taking on an opposing view or judgement | Willing and capable expert |
| **Exposure control** | To increase awareness of subjectivity: Explicitly differentiate objective data from any opinions or uncertain interpretations of data. Similarly with data availability versus data selection. | Initiator |
| **Visualisation** | To maximise comprehension and to minimise potential for misunderstandings or error: Present data or knowledge in multiple formats (e.g., tables, line-graphs, maps) | Initiator & Preparation time |
| **Explicit knowledge** | To determine rationale for expert opinion: Ask experts to justify their answers based on explicit knowledge, e.g., "I think this is a salt body because I have seen this structure hundreds of times before and it has always been salt" | Initiator |
| **Expert identification** | To ensure task matches expertise: The scope of a project may evolve with time, as such, the expertise | Facilitator, Expert & Problem-owner |

| | required may change. Check expertise matches current task requirements. | |
|---|---|---|
| **Expert profiling** | To determine how well calibrated experts are for the task: Use seed questions where possible to obtain an expert profile, especially where uncertainty judgements are required. | Expert & Facilitator / Computer |
| **Risk attitude profile** | To identify an individual's perception of risk within a task: Assess an expert's current risk attitude in the task domain(s) via specific questions on risk and benefit (see text for further details). | Facilitator / Computer |
| **Deconstruct task** | To aid the identification of underlying sources of uncertainty or major divisive factors: Ask experts to discuss the task, and then subsequently assess sub-tasks independently that arise from the discussion. Re-construct the main task from sub-task results (see text for further details). | Facilitator & Time for multiple task combinations |
| **Reword task** | To reduce misunderstandings: Explain task in multiple ways and clarify all terminology and acronyms. This is especially important where tasks involve different expert-types. | Expert & Time |
| **Forced anonymity** | Instigate anonymous opinions: Ask experts to present their opinions anonymously to the group (e.g., via computer to a message board, or via paper). Especially important if herding is suspected or known to occur within a group. | Facilitator |